\documentstyle[twoside,psfig]{article}
\begin{document}
\twocolumn[
\begin{center}
{\LARGE \bf ENTANGLEMENT 
 AND THE LINEARITY OF}\\

\vspace*{0.62cm}
{\LARGE \bf
 QUANTUM MECHANICS}\\

\vspace*{0.8cm}
{\large \bf Gernot Alber}\\

\vspace*{0.1cm}
Abteilung f\"ur Quantenphysik,
Universit\"at Ulm, D-89069 Ulm, Germany
\\(Proceedings of the X International Symposium on
Theoretical Electrical Engineering ISTET 99)
\end{center}

\vspace*{1.0cm}
]
\noindent
{\Large \bf Abstract}\\
\\
Optimal universal entanglement processes are discussed 
which entangle two quantum systems in an optimal way
for all possible initial states. It is demonstrated that
the linear character of quantum theory which enforces the
peaceful coexistence of quantum mechanics and relativity
imposes severe restrictions on the structure of the resulting
optimally entangled states.  Depending on the dimension of
the one-particle Hilbert space such a universal process
generates either a pure Bell state or
mixed entangled states. In the limit of
very large dimensions of the one-particle Hilbert space 
the von-Neumann entropy of the
optimally entangled state 
differs from the one of the
maximally mixed two-particle state 
by one bit only.
\\
\\
{\Large \bf Introduction}\\
\\
Ever since its discovery by Schr\"odinger \cite{Schr}
the existence of entanglement between
different quantum systems has been a major puzzling
aspect of quantum theory.  
If a
quantum mechanical many particle system is in an 
entangled state its characteristic physical properties are
distributed over all its
subsystems without being present in any one of them
separately. In the newly emerging science of quantum information
processing \cite{world} these puzzling aspects of quantum theory are
recognized as a potentially useful new resource which might help
to perform various tasks of practical interest
more efficiently than it is possible by any other classical means.
Prominent examples in this respect are applications
of entangled states in secret quantum key distribution (quantum
cryptography) 
and in fast quantum algorithms (quantum computing).

In view of these developments the natural question arises
whether it is possible to design universal quantum processes
which entangle two or more quantum systems in an optimal way
for all possible initial states of the separate subsystems.
Definitely, provided the initial states of the subsystems are
known it should always be possible to design a particularly tailored
quantum process which
produces any desired quantum state.
However, this becomes less obvious if one wants to design a
universal
quantum process which is independent of possibly unknown
input states and
which performs the required task for all possible input
states in the same optimal way. Which constraints are imposed
by the fundamental laws of quantum mechanics on such universal,
optimal processes?

Recently, similar questions have been studied
extensively in the context of quantum cloning
\cite{clone1,clone2,clone3,clone4,clone5}
where one aims at
copying arbitrary quantum states by a universal quantum process.
It has been known for a long time that this task
cannot be performed perfectly
due to constraints imposed by the
linear character of quantum theory \cite{Wigner,Wooters}.
According to this linear character any quantum 
process has to map the density operator of 
the initial state linearly
onto the density operator of the final state. If the relation
between the density operators of initial and final states were not
linear, one could distinguish different unravellings of
one and the same density operator physically.
This would contradict
the  basic postulate of quantum theory that the physical state
of a quantum system is described by a density operator and not
by any of its possibly inequivalent unravellings \cite{Peres}. 
This linear character of quantum theory implies, for example,
that despite their nonlocal character 
it is not possible
to use entangled states for super luminal communication
\cite{Gisin1}.
This so called no-signaling constraint of quantum theory 
enforces the peaceful coexistence of quantum mechanics
and relativity \cite{Shimony} and
imposes severe restrictions
on universal quantum processes \cite{clone5}.
In the context of
optimal quantum cloning \cite{clone1,clone2,clone3,clone4,clone5}
and the universal NOT gate \cite{Buzek} these constraints
have already been 
investigated. However, their influence on other universal quantum
processes is still widely unknown.

Motivated by the importance which entangled states play
in the context of 
quantum information processing
in the following the question is addressed
whether it is
possible to design a universal quantum process which entangles
quantum systems in an optimal way. 
How can one define such a universal, optimal entanglement process
and which restrictions are imposed by the linear character
of quantum theory? 
What is the nature of the class of resulting optimally
entangled states?
Answering these questions sheds new light onto
the basic concept of entanglement itself and onto the question
which types of entangled states can be prepared by quantum processes
in a natural way.
\\
\newpage

\noindent
{\Large \bf Optimal Entanglement by\\ Universal Quantum Processes}
\\
\\
How can one define a universal quantum process which entangles
quantum systems in an optimal way for all possible input states?

In order to put this problem into perspective let us consider
the simplest possible situation, namely a quantum process which
entangles 
two particles whose associated 
Hilbert spaces ${\cal H}_N$ have equal dimensions
of magnitude $N$.
We assume that an arbitrary, pure
input state $\rho_{in}({\bf m})$
is entangled with a known reference state
$\rho_{ref}$
by a general quantum process 
\begin{equation}
{\cal P}: \rho_{in}({\bf m})\otimes \rho_{ref} \to
\rho_{out}({\bf m})
\label{ansatz}
\end{equation}
thereby yielding the two-particle
output state $\rho_{out}({\bf m})$.
In particular, we are looking for a universal
quantum process which
is independent of the input state and 
entangles both particles in an optimal way for all possible, pure
input states $\rho_{in}({\bf m})$.

In an N-dimensional Hilbert space
an arbitrary input state can always be represented in the form
\begin{equation}
\rho_{in}({\bf m}) = \frac{1}{N} ({\bf 1} + m_{ij} {\bf A}_{i j}) 
\end{equation}
where the operators ${\bf A}_{ij}$ ($i,j=1,...,N$)
form a basis for the Lie-Algebra
of $SU_N$ \cite{Mahler}. (We adopt the usual
convention that one has to sum over all
indices which appear twice .)
Explicitly these operators can be
represented by the $N\times N$ matrices
\begin{eqnarray}
({\bf A}_{ij})^{(kl)}=
\delta_{k i} \delta_{j l} - \delta_{ij}\delta_{kl}/N
\hspace*{0.5cm}(k,l=1,...,N)
\end{eqnarray}
with $\delta_{ij}$ denoting the Kronecker delta-function.
These operators
might be viewed as
generalizations of the Pauli spin operators
${\bf \sigma}_x,{\bf \sigma}_y$ and ${\bf \sigma}_z$ 
to cases with $N > 2$ and they
fulfill the relations ${\rm Tr}\{{\bf A}_{ij}\}=0$, 
${\bf A}^{\dagger}_{ij}={\bf A}_{ji}$. 
The characteristic quantity ${\bf m}$ whose
components are denoted $m_{ij}$ ($i,j=1,..,N$) might
be viewed as a generalized Bloch vector.
For $N=2$ the operators ${\bf A}_{ij}$ are
related to the Pauli spin operators by
${\bf \sigma}_z\equiv2{\bf A}_{11}\equiv-2{\bf A}_{22}$,
${\bf \sigma}_x+i{\bf \sigma}_y
\equiv2{\bf A}_{12}$ and
${\bf \sigma}_x-i{\bf \sigma}_y\equiv2{\bf A}_{21}$.
The self-adjointness of
the density operator $\rho_{in}({\bf m})$
implies that
$m_{ij}=m_{ji}^*$. Furthermore, $\rho_{in}({\bf m})$
represents a
pure state only if ${\rm Tr}[\rho^2_{in}({\bf m})]
= {\rm Tr}[\rho_{in}({\bf m})] = 1$ which implies
the relation
$[m_{i j}m_{j i} - (m_{ii})^2/N]=N^2(1-1/N)$.
In a similar way also
the two-particle
output state of Eq.(\ref{ansatz}) 
can be expressed in terms of
these generators of $SU_N$ according to
\begin{eqnarray}
\rho_{out}({\bf m}) &=&\frac{{\bf 1}\otimes {\bf 1}}{N^2} +
\alpha^{(1)}_{ij}({\bf m}){\bf A}_{i j}\otimes {\bf 1} +
\label{rout}
\\
&&
\alpha^{(2)}_{ij}({\bf m}){\bf 1}\otimes {\bf A}_{i j} +
K_{ij rs}({\bf m}){\bf A}_{i j}\otimes {\bf A}_{r s}.\nonumber 
\end{eqnarray}

What are the basic requirements which 
an optimal, universal entanglement process ${\cal P}$ of the general
form of Eq.(\ref{ansatz}) should fulfill?
Definitely the notion of optimal entanglement is not well defined
in particular for mixed states
due to the lack of a unique measure of entanglement
\cite{measure,measure1,measure2}.
Despite these difficulties it appears natural to 
regard the following two conditions as a
minimal set of
requirements for an optimal, universal entanglement process for
two particles, namely
\begin{eqnarray}
{\rm Tr}_{2}\{\rho_{out}({\bf m})\}  &=&
{\rm Tr}_{1}\{\rho_{out}({\bf m})\} = \frac{{\bf 1}}{N},
\label{1}
\end{eqnarray}
\begin{eqnarray}
S[\rho_{out}({\bf m})] &=& -
{\rm Tr}\{\rho_{out}({\bf m}) {\rm ln}[\rho_{out}({\bf m})]\} \to
{\rm minimum}\nonumber\\
~
\label{2}
\end{eqnarray}
for all possible input states $\rho_{in}({\bf m})$.
The first condition expresses the well know property
of pure, two-particle entangled states that
they behave as  maximally mixed states as far as all
one-particle properties are concerned. 
(${\rm Tr}_{1(2)}$ denotes the trace over the state space of particle
$1 (2)$.)
The second condition states that the entangled
two-particle output state
$\rho_{out}({\bf m})$
should be as pure as possible so that the associated von-Neumann
entropy $S[\rho_{out}({\bf m})]$ is minimal. (In Eq.(\ref{2}) 
this entropy is measured in units of Boltzmann's constant.)
Together
these two requirements imply that
in the resulting quantum state $\rho_{out}({\bf m})$
the quantum information
is distributed over both particles without
being present in each one of the separate particles alone.
If one does not consider both particles together one looses
a maximum amount of information. In this sense the requirements
(\ref{1}) and (\ref{2}) characterize optimal entanglement between
both particles.
In the subsequent treatment it is demonstrated that these
two conditions 
which concentrate on the information theoretic aspects of
entanglement
characterize uniquely a universal quantum
process which 
yields entangled two-particle output states for arbitrary dimensions
of the one-particle Hilbert space ${\cal H}_N$. 

Conditions (\ref{1}) and 
(\ref{2}) imply that an optimal $\rho_{out}({\bf m})$
can always be found by
the covariant ansatz
\begin{eqnarray}
\rho_{out}({\bf R} {\bf m}) &=& U({\bf R})\otimes U({\bf R})
\rho_{out}({\bf m})
U^{\dagger}({\bf R})\otimes U^{\dagger}({\bf R})
\nonumber\\
~
\label{cov}
\end{eqnarray}
for all possible ${\bf R}\in SU_N$.
Eq.(\ref{cov}) states that the set
of possible two-particle output states $\rho_{out}({\bf m})$
forms a representation of the group
$SU_N \times SU_N$ thus ensuring that
the von-Neumann entropy is the same for all possible input states
$\rho_{in}({\bf m})$.
Thereby ${\bf R}\in SU_N$ represents the particular
unitary transformation with matrix elements $R_{ijkl}$ 
which transforms a given input state 
$\rho_{in}({\bf m})$
into an arbitrary other input state $\rho_{in}({\bf m}')$
according to the
transformation law $m'_{ij} = R_{ijkl}m_{kl}$.
In fact, as conditions (\ref{1}) and (\ref{2})
are independent of the input state,
the optimal output state can even be found by the more restrictive
invariant ansatz
$\rho_{out}({\bf R}{\bf m}) = \rho_{out}({\bf m})$ 
for all possible ${\bf R}\in SU_N$. This ansatz
is a special
case of the covariant relation of Eq.(\ref{cov}).
However, as we want to investigate universal quantum processes also in
a more general context we do not want to impose this more restrictive
invariance condition already from the very beginning.
Furthermore, as conditions (\ref{1}) and (\ref{2})
are also invariant under
permutations of the particles an optimal $\rho_{out}({\bf m})$
also has to be permutation invariant.

Apart from the covariance condition of Eq.(\ref{cov})
any quantum process also has to be compatible
with the linearity of quantum mechanics. This linearity
implies that
$\rho_{out}({\bf m})$ has to be a linear function of the generalized
Bloch vector
${\bf m}$ which characterizes the input state. Thus
the characteristic quantities
$\alpha^{(1)}_{ij}({\bf m}),\alpha^{(2)}_{ij}({\bf m})$ and
$K_{ij rs}({\bf m})$ of Eq.(\ref{rout}) all have to be linear
functions of ${\bf m}$. This linear dependence guarantees
that different unravellings of the same
input state yield the same output state 
after application of the
universal quantum process so that this process cannot
distinguish between different unravellings 
of $\rho_{in}({\bf m})$. 

Both the covariance condition of Eq.(\ref{cov}) and the linearity
constraint impose severe restrictions on 
general universal quantum processes of the form of Eq.(\ref{rout}).
\\
\\
{\Large \bf   Covariant and linear universal\\ quantum processes}
\\
\\
What is the
structure of
general covariant and linear quantum processes which result in
a two-particle output state which is invariant under permutations
of both particles?
Answering this question will yield a unified theoretical
description for a general class of universal
quantum processes which include  both universal optimal
quantum cloning and 
universal optimal entanglement as special cases.

The covariance condition of
Eq.(\ref{cov}) can be implemented easily 
by observing that
only a tensor product of the form
${\bf S}={\bf A}_{ij}\otimes {\bf A}_{ji}$
transforms as a scalar under $SU_N\times SU_N$, i.e.
${\bf U}({\bf R})\otimes {\bf U}({\bf R}) {\bf S}
{\bf U}^{\dagger}({\bf R})\otimes {\bf U}^{\dagger}({\bf R}) = {\bf S}$.
Similarly, it is straightforward to demonstrate
that only tensor products of the form
${\bf V}_{il}={\bf A}_{ij}\otimes {\bf A}_{jl}$ or 
${\bf V}^{\dagger}_{il}$ transform like generalized vectors
under $SU_N\times SU_N$, i.e.
${\bf U}({\bf R})\otimes {\bf U}({\bf R}) {\bf V}_{il}
{\bf U}^{\dagger}({\bf R})\otimes {\bf U}^{\dagger}({\bf R})
= {\bf V}_{km}R_{kmil}$.
Thus the most general two-particle
quantum process which is covariant, linear
in ${\bf m}$ and invariant under permutations of both
particles is of the form
\begin{eqnarray}
\rho_{out}({\bf m}) &=&\frac{{\bf 1}\otimes {\bf 1}}{N^2}
+
\alpha m_{i j} {\bf A}_{ij}\otimes {\bf 1} +\nonumber \\
&&\alpha m_{i j} {\bf 1}\otimes {\bf A}_{ij} +
C{\bf A}_{i j}\otimes {\bf A}_{j i} +\nonumber\\
&&
\beta m_{il} {\bf A}_{ij}\otimes {\bf A}_{jl} +
\beta m_{li}{\bf A}_{ji}\otimes {\bf A}_{lj}
\label{routcov}
\end{eqnarray}
and is characterized uniquely by the
real-valued parameters $\alpha,\beta$ and $C$.
These characteristic parameters have to be restricted to
their physical domain which is defined by the
requirement that 
$\rho_{out}({\bf m})$ is a density operator
and must have
non-negative eigen values with
${\rm Tr}[\rho_{out}({\bf m})] = 1$.
%So far Eq.(\ref{routcov}) is quite general and
%the conditions (\ref{1}) and (\ref{2}) for optimal entanglement
%have not been incorporated yet.

In order to obtain insight into the physical contents of the class
of universal, covariant and linear quantum processes
which is described by Eq.(\ref{routcov})
let us investigate the structure of $\rho_{out}({\bf m})$
more explicitly. Due to the covariance condition
we can restrict ourselves to a pure input state 
with
$m_{ij} = N \delta_{i1}\delta_{j1}$
without loss of generality.
Introducing an orthogonal
basis $\{|1\rangle,...,|N\rangle \}$ in the N-dimensional
one-particle Hilbert space ${\cal H}_N$
in which state $|1\rangle$ denotes the input state,
i.e. $\rho_{in}({\bf m}=m_{11}{\bf e}_{11}) = |1\rangle \langle 1|$,
one obtains from Eq.(\ref{routcov}) the expression
\begin{eqnarray}
\rho_{out}({\bf m}&=&m_{11}{\bf e}_{11}) =
M_{11}|11\rangle \langle 11| +\nonumber\\
&& \sum_{j=2}^{N} |jj\rangle \langle jj|
(M_{23}+C) +
\nonumber\\
&&
\sum_{j=2}^{N} \{
|1j\rangle \langle 1j| M_{12} +
|1j\rangle \langle j1| (C+\beta m_{11}) +\nonumber\\
&&
|j1\rangle \langle 1j| (C+\beta m_{11}) +
|j1\rangle \langle j1|  M_{12}
\}+\nonumber\\
&&\sum_{i<j=2}^{N} \{
|ij\rangle \langle ij| M_{23} +
|ij\rangle \langle ji| C +\nonumber\\
&&
|ji\rangle \langle ij| C +
|ji\rangle \langle ji| M_{23}
\}
\label{explicit}
\end{eqnarray}
with
\begin{eqnarray}
M_{23}&=&1/N^2-2\alpha m_{11}/N-C/N+2\beta m_{11}/N^2,\nonumber\\
M_{12}&=&M_{23}+\alpha m_{11}-2\beta m_{11}/N,\nonumber\\
M_{11}&=&1/N^2+2\alpha m_{11}
(1-1/N)+C(1-1/N)+\nonumber\\
&&2\beta m_{11}(1-1/N)^2.\nonumber
\end{eqnarray}
The non-negativity of $\rho_{out}({\bf m})$ implies
the constraints
$M_{23}\geq|C|$, $M_{12}\geq|C+\beta m_{11}|$, $M_{23}+C\geq 0$ and
$M_{11}\geq 0$.

The two-particle output states of
Eqs.(\ref{routcov}) or (\ref{explicit})
characterize all possible permutation invariant,
covariant, linear mappings.
They describe in a unified way the restrictions which are imposed
by the linearity of quantum mechanics on universal
quantum processes which treat both particles in a symmetric way.
The universality of these processes guarantees
that they fulfill any additional conditions
for all possible input states. 
The general covariance condition of Eq.(\ref{cov})
implies that
these additional conditions need not be invariant under
unitary transformations. They may very well depend on properties
of the initial input state.

As a special case of such a universal
quantum process let us consider 
optimal cloning of pure states
\cite{clone1,clone2,clone3,clone4,clone5}.
In this case one is looking for a 
mapping ${\cal P}$ of the form of Eq.(\ref{ansatz})
which fulfills the additional constraint
\begin{equation}
{\rm Tr}\{
\rho_{in}({\bf m})\otimes \rho_{in}({\bf m})
\rho_{out}({\bf m})
\} \to {\rm maximum}
\label{clone}
\end{equation}
for all possible input states $\rho_{in}({\bf m})$.
This constraint involves the input state explicitly and it is
equivalent to maximizing $M_{11}$ in Eq.(\ref{explicit}).
Physically speaking this condition imposes the constraint that
the output state $\rho_{out}({\bf m})$ should be as close as possible
to the ideally cloned state
$\rho_{in}({\bf m})\otimes \rho_{in}({\bf m})$.
It is straightforward to work out the optimal parameters which
satisfy Eq.(\ref{clone}), namely
$2\alpha m_{11} = (N+2)/[N (N+1)], \beta m_{11}=1/[2N+2] , C = 0$.
Inserting these parameters into Eq.(\ref{explicit}) one realizes that
optimal cloning can be achieved only imperfectly with a probability
of $P_{11} \equiv M_{11} = 2/(N + 1) < 1$. With a probability of
$1-P_{11} = (N-1)/(N+1)$
in this process also an unavoidable maximally mixed
state is generated which involves
all possible Bell states of the form
$|\psi_{1j}\rangle^{(+)}
= (|1 j \rangle + |j 1\rangle)/\sqrt{2}$ with
equal probabilities.
Thereby state $|j\rangle$ can be any of the $(N-1)$ basis
states which are orthogonal to the pure input state
$\rho_{in}({\bf m}=m_{11}{\bf e}_{11})$.
Thus the two-particle output state
of the universal, optimal quantum cloning process
is given by
\begin{eqnarray}
\rho_{out}({\bf m}&=&m_{11}{\bf e}_{11}) = P_{11}|11\rangle \langle 11|
+\nonumber\\
&&  \frac{(1 - P_{11})}{N-1}
\sum_{j=2}^N|\psi_{1j}\rangle^{(+)} ~^{(+)}\langle \psi_{1j}|.
\end{eqnarray}
\\
\\
{\Large \bf 
Nature of the universal, optimally entangled
two-particle states}
\\
\\
What is the nature of the entangled states which are produced
by the optimal entanglement process characterized by
the covariant and linear map of Eq.(\ref{explicit})
and by conditions
(\ref{1}) and (\ref{2})?

Let us first of all
determine the values of the characteristic parameters
$\alpha, \beta$ and $C$ 
of this universal, optimal entanglement process.
Condition (\ref{1})
implies that $\alpha =0$. Minimizing the von-Neumann entropy
$S[\rho_{out}({\bf m})]$
implies that we have to determine the remaining parameters
$\beta$ and $C$ in such a way that the number of eigen values
of magnitude zero is as large as possible.
The physical region of the two remaining parameters
\begin{figure}[h]
\centerline{\psfig{figure=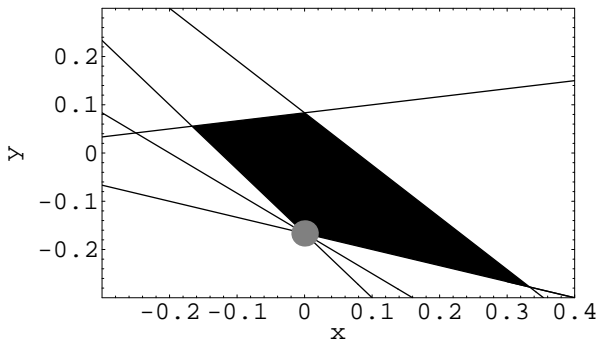,width=8.6cm,clip=}}
\caption{\small Schematic representation of the physical 
region  of the parameters
$y=C$ and $x=\beta m_{11}$ (black) for $\alpha=0$ and $N=3$.
It is determined by the requirement that $\rho_{out}({\bf m})$
has to be non-negative.
Each straight line indicates the parameter values for which
one of the eigen values of $\rho_{out}({\bf m})$ is zero.
The grey
dot indicates the condition for optimal, universal entanglement.
It is the only point in which three types of eigen values
of $\rho_{out}({\bf m})$
are zero
simultaneously.}
\label{fig1}
\end{figure}
$C$ and $\beta m_{11}$ is indicated in Fig. 1 by the black area.
From Fig. 1 it is straightforward to
show that the condition of minimal entropy
is fulfilled for $\beta=0$ and
$C = -1/[N(N - 1)]$. This implies that 
$\rho_{out}({\bf m})$ transforms indeed as a scalar
under $SU_N\times SU_N$ as we have already anticipated earlier.
Thus the two-particle output state
which is produced by the optimal, universal entanglement process 
is independent of the input state $\rho_{in}({\bf m})$ and
is given by
\begin{eqnarray}
\rho_{out}({\bf m}) &=&\frac{2!}{N ( N - 1)}\sum_{i<j=1}^N
|\psi_{ij}\rangle^{(-)}~^{(-)}\langle \psi_{ij}|.
\nonumber\\
~\label{ent}
\end{eqnarray}
In general, $\rho_{out}({\bf m})$ is a maximally disordered mixture
of all possible anti-symmetric Bell states
\begin{eqnarray}
|\psi_{ij}\rangle^{(-)} &=& \frac{1}{\sqrt{2}}
(|ij\rangle - | ji \rangle)
\end{eqnarray}
which can
be formed by two possible basis states
$|i\rangle$ and $|j\rangle$ of the N-dimensional one-particle
Hilbert space ${\cal H}_N$.
The number of these anti-symmetric Bell states is given by
$[N(N-1)/2!]={N \choose 2}$. 
It is interesting to realize that it is only the anti-symmetric
Bell states which appear in this optimal, universal
entanglement process. This is understandable from the fact that
these Bell states are the only ones which are invariant under
arbitrary unitary transformations. This invariance property
guarantees that one obtains entangled output states for all
possible input states so that the resulting entanglement
process is universal. 
The other three two-particle
Bell states, namely
\begin{eqnarray}
|\psi_{ij}\rangle^{(+)} &=&\frac{1}{\sqrt{2}}
(|ij\rangle + | ji\rangle),
\nonumber\\
|\Phi_{ij}\rangle^{(\pm)} &=&\frac{1}{\sqrt{2}}
(|ii\rangle  \pm |  jj\rangle),
\end{eqnarray}
do not have this invariance property.
If they appeared in the two-particle output state, it would always
be possible to find a particular input state which produces a
separable, non-entangled output state. Thus such a quantum process
would not fulfill the universality requirement.

For the case of universal optimal entanglement of a qubit, i.e. for
$N=2$, there is only one possible anti-symmetric Bell state, namely
$|\psi_{12}\rangle^{(-)}$.
Thus in this particular case 
the universal entanglement process of Eq.(\ref{ent})
produces the pure two-particle output state
$\rho_{out}({\bf m}) =
|\psi_{12}\rangle^{(-)}~^{(-)}\langle \psi_{12}|$
which is known to violate Bell inequalities maximally \cite{Peres}.
For all higher values of
the dimension of the Hilbert space ${\cal H}_N$ the two-particle
output state is mixed.
Nevertheless according to condition (\ref{2})
the von-Neumann entropy of all possible
output states is always as small as possible
within the linearity constraints
imposed by quantum theory.
Furthermore, it is straightforward
to show that all output states are not separable 
as their partial transposes have at least one
negative eigen value \cite{Peres1}
of magnitude $\lambda = -1/[N(N-1)] < 0$.

How do these optimal, universal
two-particle output states behave for high values
of the dimension of the one-particle Hilbert space
${\cal H}_N$?
In general the von-Neumann entropy of the two-particle output
state is given by
\begin{equation}
S[\rho_{out}({\bf m})] = {\rm ln} {N \choose 2} =
{\rm ln}[N(N-1)] - {\rm ln}[2!].
\end{equation}
For $N\gg 1$ this entropy approaches the value\\
$S[\rho_{out}({\bf m})] \to {\rm ln}[N^2] - {\rm ln}[2!]$.
Thereby ${\rm ln}[N^2]$ is the entropy of the maximally disordered
two-particle state $\rho_{max} = {\bf 1}\otimes {\bf 1}/N^2$.
Thus, in the limit
of large dimensions of the Hilbert space ${\cal H}_N$
the entropies of $\rho_{max}$ and of 
$\rho_{out}({\bf m})$ differ by one bit only .
This shows that in the limit of large $N$
these universal, optimally entangled states
are located very close to the 
maximally mixed state $\rho_{max}$ 
from the information theoretic point of view.
They are very fragile with respect to any disturbances.
Loosing one bit of information only changes them to a maximally
mixed state $\rho_{max}$.
Nevertheless, it is worth pointing out that this does not
necessarily
imply that these states are also close to $\rho_{max}$ in state
space. In order to characterize the distance of a mixed
quantum state $\rho$
from the maximally mixed one in state space one
usually decomposes $\rho$ according to
\begin{eqnarray}
\rho &=& (1 - \epsilon){\bf 1}/d + \epsilon \rho_1
\end{eqnarray}
with a suitably chosen density operator $\rho_1$ (with\\
${\rm Tr}[\rho_1] = 1$) and with $d$ denoting the dimension of the
relevant Hilbert space.
The quantity $\epsilon$ might be considered as characterizing
the separation of $\rho$ from the maximally mixed state.
Mixed states which are close to the maximally mixed one in the sense
that $0\leq \epsilon \ll 1$ are of particular interest for
quantum information processing in high-temperature nuclear magnetic
resonance \cite{nuc1,nuc2,nuc3}.
In this context Braunstein et al. \cite{Schack}
have shown recently that in systems consisting
of n-qubits with $d=2^n$ one can always find a sufficiently small
neighborhood around the maximally mixed state with
$\epsilon = O(4^{-n})$ within which all states are separable.
In view of this result it is of interest to work out also the
distance of the
universal, optimally entangled states of Eq.(\ref{ent})
from the maximally mixed state $\rho_{max}$ in state space.
As many of the possible $N^2$ eigen values of $\rho_{out}({\bf m})$
are zero these states
are characterized by $\epsilon = 1$. Thus, despite their
closeness to $\rho_{max}$ as far as the von-Neumann entropy
is concerned, these latter states are well separated
from $\rho_{max}$ in state space for all possible values of the
dimension of the Hilbert space ${\cal H}_N$.
\\
\\
{\Large \bf Acknowledgments}\\
\\
This work is supported by the DFG within the SPP
`Quanteninformationsverarbeitung' and by the ESF programme
`Quantum Information Theory and Quantum Computation'.
Stimulating discussions with N. Gisin are acknowledged.

\end{document}